\documentclass[aps,pra,reprint,amsmath,amssymb,superscriptaddress]{revtex4-1}

\usepackage{graphicx}
\usepackage{bm}
\usepackage{dcolumn}
\usepackage{gensymb}
\usepackage[utf8]{inputenc}
\usepackage{float}

\usepackage{url,hyperref,breakurl}

\usepackage{array,mathtools,amssymb,booktabs}
\newcolumntype{C}{>{$}c<{$}}
\AtBeginDocument{
\heavyrulewidth=.08em
\lightrulewidth=.05em
\cmidrulewidth=.03em
\belowrulesep=.65ex
\belowbottomsep=0pt
\aboverulesep=.4ex
\abovetopsep=0pt
\cmidrulesep=\doublerulesep
\cmidrulekern=.5em
\defaultaddspace=.5em
}

\usepackage{hyperref}
\newcommand{\tref}[1]{Table~\ref{#1}}
\newcommand{\eref}[1]{Eq.~(\ref{#1})}

\begin{document}

\title{Calculation of energies and hyperfine structure constants of $^{233}$U$^{+}$ and $^{233}$U}

\author{S. G. Porsev}
\affiliation{Department of Physics and Astronomy, University of Delaware, Newark, Delaware 19716, USA}
\author{C. Cheung}
\affiliation{Department of Physics and Astronomy, University of Delaware, Newark, Delaware 19716, USA}
\author{M. S. Safronova}
\affiliation{Department of Physics and Astronomy, University of Delaware, Newark, Delaware 19716, USA}
\affiliation{Joint Quantum Institute, National Institute of Standards and Technology and the University of Maryland,
College Park, Maryland 20742, USA}

\begin{abstract}
We carried out calculations of the energies and magnetic dipole hyperfine-structure constants of the low-lying states
of $^{233}$U$^{+}$ and $^{233}$U using two different approaches. With six valence electrons and a very heavy core, uranium represents a major
challenge for precision atomic theory even using large-scale computational resources. The first approach combines configuration interaction (CI)
with a method allowing us to include core-valence correlations to all orders of the perturbation theory over residual Coulomb
interaction. The second approach is a pure CI method which allows the use of different initial approximations. We present a detailed analysis
of all calculated properties and discuss the advantages and disadvantages of each of these methods.
We report a preliminary value of the U nuclear magnetic moment and outline the need for further experiments.
\end{abstract}

\date{\today}

\maketitle

% =====================
\section{Introduction}
% =====================
Experimental measurements of hyperfine structure (hfs) together with theoretical calculations can be used
to determine nuclear moments. Recently, we
calculated the hfs constants $A$ and $B$ for a heavy ion, $^{229}$Th$^{3+}$, ~\cite{PorSafKoz21}. Combining these values with the experimental
results, we extracted the magnetic dipole and electric quadrupole nuclear moments with high accuracy. The electronic structure of Th$^{3+}$, which has a single valence electron, allowed for the most precise calculation for an actinide ion. The 229 isotope of Th has a very small transition energy ($\sim 8$ eV) between the ground and first excited nuclear states.
Due to such a unique feature, this nuclear transition was proposed for the design  of a new type of optical clock (see review~\cite{PeiSchSaf21}
and references therein).

The Th ions were successfully trapped, and further precision laser spectroscopic investigation of the hyperfine structure is planned~\cite{PeiSchSaf21}. Thorium isomer energy can be measured using the decay of the isomer state in neutral Th via internal conversion \cite{SeiWenBil19}.
In these experiments, $^{229}$Th isomer is produced from the $\alpha$ decay of $^{233}$U. As a result, one can investigate the
hyperfine-structure constants of $^{233}$U$^+$ in the same set of experiments. Such an experimental study was carried out in Ref.~\cite{Thi21}
and the hfs constant $A$ of the first excited state $5f^3 6d 7s\,\,^6\!L_{11/2}$ was found with about 2\% uncertainty to be $-301(6)$ MHz.
Therefore, an accurate calculation of this constant can potentially lead to improving the current value of the nuclear magnetic moment $\mu = 0.59(5)\,\mu_N$~\cite{Sto05} (where $\mu_N$ is the nuclear magneton), known with only 8.5\% accuracy.

Motivated by these and future experiments, we carried out calculations of the energies and magnetic dipole hfs constants $A$ of the low-lying
states of $^{233}$U$^+$ and $^{233}$U. The U$^+$ ions and neutral U are                                                                                                                                                                                                                                                                                                                                                                                                                                                                                                                                                                                                                                                                                                                                                                                                                                                                                                                                                                                                                                                                                                                                                                                                                                                                                                                                                                                                                                                                                                                                                                                                                                                                                                                                                                                                                                                                                                                                                                                                  very challenging systems for precision atomic calculations. First, these are heavy atomic systems with the nuclear charge $Z=90$. Second, the main configurations of the ground states of U$^+$  and U are $5f^3\,6d7s$ and $5f^3\,6d 7s^2$, respectively.
These five or six electrons can be considered as the valence electrons, while all the rest can be treated as the core. Numerous problems that occur in calculating the hfs constants for neutral uranium were discussed in the recent paper~\cite{Sav20}.

An accurate treatment of both valence-valence and core-valence correlations in actinides with many valence electrons is a very challenging
task that is currently unsolved. In this paper, we explore its possible solutions, using two different methods. The first one is a combination
of the configuration-interaction (CI) method with a method allowing us to include core-valence correlations in the second order or all orders
of the perturbation theory over residual Coulomb interaction ~\cite{DzuFlaKoz96,SafKozJoh09}. This method proved to be very efficient for calculating energies of the low-lying levels of both the singly charged and neutral uranium. The transition energies were reproduced with an accuracy of several tens of cm$^{-1}$. This is the most accurate calculation carried out so far in such a system to the best of our knowledge.
However, we found unexpected difficulties when computing  hfs constants of the excited states of U$^+$: unusually large core-valence correlation corrections to the expectation values of the magnetic dipole hyperfine operator.
Performing a full-scale calculation in the framework of the pure CI method with a different initial approximation allowed us to circumvent this problem. In this method, we consider only valence-valence correlations, while core-valence correlations are not taken into account explicitly. The latter can be considered as a drawback of this method, but on the other hand,
we have the freedom to choose an optimal initial approximation, obtained from the Dirac-Hartree-Fock (DHF) self-consistency procedure. The advantages and disadvantages of these methods and the results obtained within each of them will be discussed in detail in the following sections.
% ========================================================
\section {CI+MBPT and CI+all-order methods of calculation}
% ========================================================
We consider U$^+$ and U as the atomic systems with a $[1s^2, ..., 6p^6]$ core and five and six valence electrons above it, respectively.
The initial DHF self-consistency procedure included the Breit interaction and was done for the core electrons, such potential is usually
denoted $V^{N-M}$, where $N$ is the total number of electrons and $M$ is the number of valence electrons. The advantage of such potential
is the easiest formulation of the perturbation theory and coupled-cluster (all-order) approaches without the appearance of the large so-called
subtraction diagrams (see detailed discussion in Ref.~\cite{2016Pb}).

The only successful attempt to use a different starting potential with a CI+all-order method, when the initial approximation
did not correspond to the self-consistent field of the core,
%that implicitly takes into account core-valence correlation
was for Pb, where the $V^{N-2}$ starting potential was used to treat the system with four valence electrons~\cite{2016Pb}.
However, Pb had a special case of a closed $6s^2$ shell, and no such analog potential can be constructed for U and U$^+$.

In U$^+$ and U, the only potential with which we can use the CI+all-order method is $V^{N-5(6)}$ for U$^+$ and U, respectively.
However, increasing $M$ leads to degrading quality of the one-electron orbitals and a much larger number of configurations that have
to be included in CI. The $5f$, $7s$, $6d$, and $7p$ orbitals were constructed in such a frozen-core potential.
The remaining virtual orbitals were formed using 40 basis set B-spline orbitals.
The basis set included partial waves with the orbital quantum number up to $l= 6$.
Quantum electrodynamic (QED) corrections were also included following Refs.~\cite{ShaTupYer13,TupKozSaf16}.

In an approach combining CI and a method allowing us to include core-valence correlations~\cite{DzuFlaKoz96,SafKozJoh09}, the wave functions and energy levels of the valence electrons were found by solving the multiparticle relativistic equation~\cite{DzuFlaKoz96},
\begin{equation}
H_{\rm eff}(E_n) \Phi_n = E_n \Phi_n,
\label{Heff}
\end{equation}
where the effective Hamiltonian is defined as
\begin{equation}
H_{\rm eff}(E) = H_{\rm FC} + \Sigma(E),
\label{Heff1}
\end{equation}
with $H_{\rm FC}$ being the Hamiltonian in the frozen-core approximation.
The energy-dependent operator $\Sigma(E)$ accounts for virtual excitations of the core electrons.
We constructed it in two ways: using (i) the second-order many-body perturbation theory (MBPT) over residual Coulomb interaction~\cite{DzuFlaKoz96}
and (ii) the linearized coupled-cluster single-double method~\cite{Koz04,SafKozJoh09}.
In the following, we refer to these approaches as the CI+MBPT and CI+all-order methods.

To check the convergence of the CI, we have performed several calculations, sequentially increasing the size of the configuration space.
For U$^+$, the smallest set of the configurations was constructed by including the single and double excitations from the main configurations
of the low-lying states to the shells up to $13s,13p,13d,13f$, and $13g$ (we designate it as $[13spdfg]$). We have identified seven configurations
the weight (in probability) of which exceeded 3\% for all states of interest to be
$5f^3 7s^2 6d$, $5f^3 7s 6d$, $5f^3 7s 8s$, $5f^3 6d 8s$, $5f^3 7s 7d$, $5f^2 7s 6d 6f$, and $5f^2 7s^2 6f$. All subsequent (larger) sets of configurations were constructed by allowing single and double excitations from these seven configurations.
The largest set of configurations, for which the saturation was practically reached, included the single and double excitations
to $[20spd17f13g]$ orbitals.

The same approach was used for constructing sets of configurations for neutral U. The single and double excitations were allowed from the
configurations the weight of which exceeded 3\%. For U, the largest set of configurations $[20spd17f13g]$ included $131 \times 10^6$ determinants.
%-------------------------
\subsection{Energy levels}
\label{energies}
%-------------------------
The excitation energies of the lowest-lying states of U$^{+}$ and U obtained in different approximations are listed (in cm$^{-1}$)
in~\tref{Energ}.
%The excitation energies  are counted from the ground state.
The assignments of the U$^{+}$ and U levels are from Ref.~\cite{BlaWyaVer94} and the NIST database~\cite{RalKraRea11}, respectively.
%#####################################################################################################################################
\begin{table*}[tp]
\caption{The excitation energies (in cm$^{-1}$) calculated
in the CI+MBPT and CI+all-order approximations, are presented. The CI+all-order values are listed for different sets of configurations.
The QED corrections are given in the column labeled ``QED''. The final values, given in the column labeled ``Final'', are found as the sum of the CI+all-order values obtained for the $[20spd17f13g]$ configuration set and the QED corrections.
The experimental values are given in the last two columns.}
\label{Energ}%
\begin{ruledtabular}
\begin{tabular}{clccccccccc}
&\multicolumn{1}{c}{Level} & \multicolumn{1}{c}{CI+MBPT} & \multicolumn{5}{c}{CI+all-order}
& \multicolumn{1}{c}{Final} & \multicolumn{2}{c}{Experiment} \\
&\multicolumn{1}{c}{} & \multicolumn{1}{c}{$[13spdfg]$} & \multicolumn{1}{c}{$[13spdfg]$} & \multicolumn{1}{c}{$[15spdf13g]$}
& \multicolumn{1}{c}{$[17spdf13g]$} & \multicolumn{1}{c}{$[20spd17f13g]$} & \multicolumn{1}{c}{QED} & \multicolumn{1}{c}{}
& \multicolumn{1}{c}{Ref.~\cite{BlaWyaVer94}} & \multicolumn{1}{c}{Ref.~\cite{RalKraRea11}}\\
\hline      \\[-0.6pc]
U$^+$ & $5f^3 7s^2 \,\,^4\!I_{9/2}$  &    0    &    0    &    0    &    0    &    0    &    0    &   0    &    0   &        \\[0.1pc]
      & $5f^3 6d 7s\,\,^6\!L_{11/2}$ &  1411   &   497   &   498   &   409   &   450   &  -175   &  274   &   289  &        \\[0.1pc]
      & $5f^3 6d 7s\,\,^6\!K_{9/2}$  &  1896   &  1088   &  1088   &  1003   &  1033   &  -170   &  862   &   915  &        \\[0.1pc]
      & $5f^3 6d 7s\,\,^6\!L_{13/2}$ &  3008   &  2006   &  2005   &  1907   &  1934   &  -188   & 1746   &  1749  &        \\[0.1pc]
      & $5f^3 6d 7s\,\,^6\!K_{11/2}$ &  3382   &  2471   &  2470   &  2377   &  2396   &  -180   & 2215   &  2295  &        \\[0.5pc]

 U    & $5f^3 6d 7s^2 \,\,^5\!L_6$   &   0     &    0    &    0    &    0    &    0    &    0    &    0   &        &    0   \\[0.1pc]
      & $5f^3 6d 7s^2 \,\,^5\!L_7$   &  4420   &  3772   &  3779   &  3775   &  3784  &     8    &  3792  &        &  3801  \\[0.1pc]
      & $5f^3 6d 7s^2 \,\,^5\!K_6$   &  4757   &  4221   &  4231   &  4225   &  4225  &    10    &  4235  &        &  4276
\end{tabular}
\end{ruledtabular}
\end{table*}
%#####################################################################################################################################

In the third column, we present the CI+MBPT values. They were obtained only for the $[13spdfg]$ set of configurations to demonstrate
the size of the higher-order core-valence corrections. In columns 4-7, we present the CI+all-order values for different sets of configurations.
We find that the transition energies are not very sensitive to the size of the configuration space.
%interaction set of functions.
They change only slightly (by several tens of cm$^{-1}$) when the set of configurations
is increased from $[13spdfg]$ to $[20spd17f13g]$. Following Ref.~\cite{TupKozSaf16}, we calculated
the QED corrections to the energy levels. They are listed in the column labeled ``QED''.
The final values, given in the column labeled ``Final'', are the sum of the CI+all-order values obtained for the $[20spd17f13g]$ configuration set and the QED correction. We see excellent agreement between the theoretical and experimental values. Even for the small transition energy between the ground state
of U$^+$ and the first excited state $^6\!L_{11/2}$, which is 289 cm$^{-1}$, the difference between the theory and experiment is
only at the level of 5\%. We also note that the main configuration of the ground state is $5f^3 7s^2$ while the main configuration of the
excited states is $5f^3 6d 7s$. Thus, we reproduce equally well the energies of the states belonging to the different configurations.
%=========================================================
\subsection{Magnetic dipole hyperfine structure constants}
\label{hfs}
%=========================================================
The hfs coupling due to nuclear multipole moments may be represented as a scalar product of
two tensors of rank $k$:
\begin{equation*}
H_{\mathrm{hfs}}=\underset{k}{\sum}\left( \mathbf{N}^{(k)}\cdot \mathbf{T}^{(k)}\right) ,
\end{equation*}
where $\mathbf{N}^{(k)}$ and $\mathbf{T}^{\left( k\right) }$ act in the nuclear and electronic coordinate space, respectively.
% Using this expression we write the matrix element (ME) of the operator $H_{\mathrm{hfs}}$ as
% \begin{eqnarray}
% &&\langle \gamma'IJ'FM_{F}|H_{\mathrm{hfs}}|\gamma IJ FM_{F} \rangle = (-1)^{I+J'+F} \nonumber \\
% &\times& \underset{k=1}{\sum }\langle I ||N^{(k)}|| I\rangle
% \langle \gamma'J' ||T^{(k)}||\gamma J \rangle
% \left\{
% \begin{tabular}{lll}
% $I$ & $I$  & $k$ \\
% $J$ & $J'$ & $F$%
% \end{tabular}
% \right\} .
% \label{Hhfs}
% \end{eqnarray}
% Here $I$ is the nuclear spin, $J$ is the total angular momentum of the electrons,
% ${\bf F} = {\bf I} + {\bf J}$, and $\gamma$ encapsulates all other electronic quantum numbers.
In the following, we restrict ourselves to the first term in the sum over $k$, considering only the interaction of the magnetic dipole
nuclear moment with the electrons, i.e.,
\begin{equation*}
H_{\rm hfs} \approx \mathbf{N}^{(1)} \cdot \mathbf{T}^{(1)} \equiv \mathbf{N} \cdot \mathbf{T}.
\end{equation*}

We define $\mathbf{N}$ in a dimensionless form, expressing it through the nuclear magnetic dipole moment $\boldsymbol \mu$ as
\begin{eqnarray*}
\mathbf{N} = \boldsymbol{\mu}/\mu_{N} .
\end{eqnarray*}

The operator $\mathbf{T}$ is the sum of the one-particle operators
\begin{eqnarray*}
\mathbf{T} = \sum_{i=1}^{N} \mathbf{T}_i .
\end{eqnarray*}
%where $N$ is the total number of the electrons.

Assuming the nucleus to be a charged ball of uniform
magnetization with radius $R$, the expression for one-particle electronic tensor $\mathbf{T}_i$ is given
(in atomic units $\hbar=|e|=m=1, \, c \approx 137$) by
\begin{eqnarray*}
{\bf T}_i = \frac{ {\bf r}_i \times \boldsymbol{\alpha}_i}{c\,r_{i\,>}^3}\, \mu_N,
\end{eqnarray*}
where $\boldsymbol{\alpha}_i$ is the Dirac matrix, $r_i$ is the radial position of the $i$th electron,
${\bf r}_i \times \boldsymbol{\alpha}_i$ is the vector product of ${\bf r}_i$ and $\boldsymbol{\alpha}_i$,
and
\begin{eqnarray}
 r_{i\,>} \equiv
\left\{
\begin{array}{r}
r_i,\,\,{\rm if}\,\, r_i \geq R,\\
R,\,\,{\rm if}\,\, R > r_i .
\end{array}
\right.
\end{eqnarray}

The formula connecting the hfs constant $A$ of an atomic state $\left| \gamma J\right\rangle$ with the reduced matrix element
$\langle \gamma J ||T|| \gamma J \rangle$ of the electronic tensor $\mathbf{T}$ is
\begin{eqnarray*}
A = \frac{g_N}{\sqrt{J(J+1)(2J+1)}} \langle \gamma J ||T|| \gamma J \rangle ,
\end{eqnarray*}
where $\gamma$ encapsulates all other quantum numbers, $g_N = \mu/(\mu_N\,I)$, and $I$ is the nuclear spin.

We determine the hfs constants $A$ for the low-lying states of the neutral and singly ionized uranium.
For these calculations, we use the value of the nuclear magnetic moment of $^{233}$U, $\mu = 0.59(5)\, \mu_N$~\cite{Sto05}
($I=5/2$). We note that its uncertainty, 8.5\%, is large.

The results of several computations demonstrating the size of various contributions are summarized in Table~\ref{hfs}.
In the third column, we present the results obtained at the CI+MBPT stage, when the core-valence correlations are included in the second order
of the perturbation theory. The  CI+all-order results that include higher-order core-valence correlations are obtained for the largest $[20spd17f13g]$ set of configurations. These values are given in the fourth column.
% ###################################################################################################################
\begin{table*}[tbp]
\caption{Contributions to the magnetic dipole hfs constants $A$ (in MHz), calculated with $\mu = 0.59\,\mu_N$, are presented.
The CI+MBPT and CI+all-order values are listed in the columns labeled ``CI+MBPT'' and ``CI+All'', respectively. The RPA corrections to the hfs operator are listed in the column labeled ``RPA''.
The core Brueckner ($\sigma$), structural radiation (SR), two-particle (2P), QED, and normalization (Norm.) corrections
are given in respective columns and summarized in the column labeled ``Tot. corr.''.
The values in the column labeled ``Final'' are obtained as the sum of the ``CI+All'' + ``RPA'' + ``Tot. corr.'' values.
Uncertainties are given in parentheses.}
\label{hfs}%
\begin{ruledtabular}
\begin{tabular}{clccrrrrrrccc}
      &                               &CI+MBPT & CI+All &  RPA  &$\sigma$&  SR   &  2P   & QED  & Norm. & Tot. corr.&  Final & Experim.           \\
\hline \\ [-0.5pc]
U$^+$&$A(5f^3\, 7s^2\;\;^4\!I_{9/2})$ &   150  &   140  &   -2  &    3   &   31  &   -4  &  -3  &  -29  &  -1   &$137(10)$&                   \\[0.1pc]
     &$A(5f^3\, 6d7s\;\;^6\!L_{11/2})$&  -128  &  -120  &  -26  &   27   &   37  & -162  &   6  &   43  & -50   &  -196  & -$301(6)^{\rm a}$  \\[0.1pc]
     &$A(5f^3\, 6d7s\;\;^6\!K_{9/2})$ &  -173  &  -157  &   -7  &   30   &   26  & -193  &  11  &   52  & -73   &  -237  &                    \\[0.1pc]
     &$A(5f^3\, 6d7s\;\;^6\!L_{13/2})$&   369  &   344  &   33  &  -16   &   34  &  135  &  -2  &  -95  &  56   &   432  &                    \\[0.1pc]
     &$A(5f^3\, 6d7s\;\;^6\!K_{11/2})$&   399  &   369  &   46  &  -19   &   27  &  145  &  -3  & -102  &  48   &   463  &                    \\[0.1pc]
\hline \\ [-0.5pc]
 U   &$A(5f^3\, 6d 7s^2\;\;^5\!L_6)$  &   125  &   121  &    5  &    4   &   37  &   -3  &  0.9 &  -30  &   8   &   134  &  $131.56(10)^{\rm b}$ \\[0.1pc]
     &$A(5f^3\, 6d 7s^2\;\;^5\!L_7)$  &   127  &   122  &  -18  &    1   &   28  &    3  &  0.4 &  -24  &   9   &   113  &  $122^{\rm c}$   \\[0.1pc]
     &$A(5f^3\, 6d 7s^2\;\;^5\!K_6)$  &   129  &   124  &  -10  &    1   &   23  &    3  &  0.4 &  -25  &   2   &   116  &  $128^{\rm c}$ %\\[0.4pc]
\end{tabular}
\begin{flushleft}
$^{\rm a}$Reference~\cite{Thi21}; \\
$^{\rm b}$Reference~\cite{GanZemKol97}; \\
$^{\rm c}$Obtained using the results of Ref.~\cite{AvrGinPet94} and the ratio $A(^{233}{\rm U})/A(^{235}{\rm U}) \approx -2.1722$
(see \eref{ratio} and the discussion after it).
\end{flushleft}
\end{ruledtabular}
\end{table*}
% ###################################################################################################################

\begin{figure}[h!]
\begin{center}
\includegraphics[width=\linewidth]{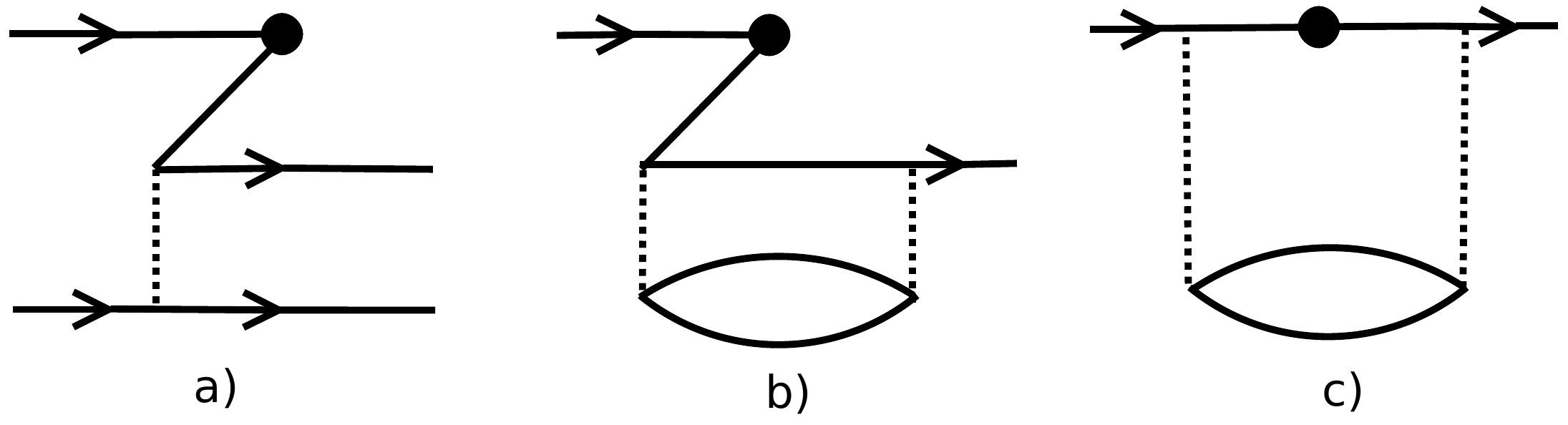}
%\includegraphics[scale=0.4]{Diagr1.pdf}
%\vspace{-0.7cm}
\caption{\label{Diagr} Representative diagrams for (a) the two-particle correction, (b) the core Brueckner correction, and
(c) the structural radiation correction. A filled circle stands for the hfs operator and a dotted line stands for the Coulomb interaction.}
\end{center}
\end{figure}

The correlation corrections to the hfs expectation values arise from the correlation corrections to the wave functions and the corrections to
the hfs operator. The most important and (often) the largest one is the random-phase approximation (RPA) correction, which is calculated
in all orders of the perturbation theory and given separately in the table in the column labeled ``RPA''.
We also took into account other corrections to the hfs operator: the two-particle (2P) and core Brueckner ($\sigma$)~\cite{DzuKozPor98}, and
the structural radiation (SR)~\cite{DzuFlaSil87,BlaJohLiu89}, QED, and normalization (Norm.) corrections. They are listed in the respective columns.
Representative diagrams for the 2P, $\sigma$, and SR corrections are shown in Fig.~\ref{Diagr}.
%------------------------------------------------------------------------------------------------------------------------------------------------
%\onecolumngrid\
%\twocolumngrid\
%------------------------------------------------------------------------------------------------------------------------------------------------

Thus, the 2P correction appears already in the first order of the perturbation theory, while the $\sigma$ and SR corrections appear in
the second order of MBPT. As illustrated in \tref{hfs}, the 2P corrections
are very large for all excited states for U$^+$. For instance, this correction is more than 100\% of the ``CI+All'' value for
the $^6\!L_{11/2}$ and $^6\!K_{9/2}$ states, which is very unusual. We had not observed it earlier for any other system.
%to the best of our knowledge.
On the other hand, the CI+all-order method has not been applied before to compute hfs for such a complicated system. The calculations of the U$^+$ properties were practically not carried out before, and specific features of these particular states are little studied. Our finding raises the question of the applicability of the perturbation theory for calculating this correction.
We can expect that higher-order corrections will play an essential role and can significantly change the hfs constants values for the excited states. Therefore, the current results obtained for these states have low accuracy, estimated to be about 50\%.

For the ground state $5f^3\, 7s^2\;^4\!I_{9/2}$, in contrast, the situation is much better. The 2P correction is small. We assume
that it can be due to the $7s$ shell being closed and that it does not contribute to the hfs constant. Besides that, all corrections beyond RPA tend to cancel each other. As a result, the total correction, labeled in \tref{hfs} as ``Tot. corr.'', and found as the sum of $\sigma$, SR, 2P, QED, and normalization corrections, is small ($\sim$ 0.7\%).
%This is common and usually expected to happen.
Estimating the uncertainty of $A(^4\!I_{9/2})$ as the difference between the ``CI+All'' and ``CI+MBPT'' values, we arrive at
$A(^4\!I_{9/2}) = 137(10)$ MHz.

For the neutral uranium, there is good agreement (about 2\%) for the hfs constant of the ground state $A(^5\!L_6)$ while for the excited
states $^5\!L_7$ and $^5\!K_6$ the agreement is at the level of 10\%. Same as for the ground state of U$^+$, the corrections
to the hfs constants are not very large and tend to cancel each other. The 2P corrections are small, which again can be explained by the presence of the closed $7s$ shell.

We note that the experimental values $A(^5\!L_7)=-56.31(12)\,{\rm MHz}$ and $A(^5\!K_6)=-59.13(45)\,{\rm MHz}$ are known with high
accuracy for $^{235}$U~\cite{AvrGinPet94}, but they were not measured for $^{233}$U. In Ref.~\cite{GanZemKol97} the magnetic dipole hfs constants
of the ground state $(5f^3\, 6d 7s^2\,^5\!L_6)$ and two even-parity states $(5f^3\, 6d 7s 7p\,\,^7\!M_7, E=16900\, {\rm cm}^{-1})$ and
$(5f^3\, 6d 7s 7p\,\,^7\!L_6, E=17362\, {\rm cm}^{-1})$ were measured for both $^{233}$U and $^{235}$U.

Their ratios were found in Ref.~\cite{GanZemKol97} to be
\begin{eqnarray}
A_{^5\!L_7}(^{233}{\rm U})/A_{^5\!L_7}(^{235}{\rm U}) &=& -2.1656(16), \nonumber \\
A_{^7\!M_7}(^{233}{\rm U})/A_{^7\!M_7}(^{235}{\rm U}) &=& -2.1790(10), \nonumber \\
A_{^7\!L_6}(^{233}{\rm U})/A_{^7\!L_6}(^{235}{\rm U}) &=& -2.172(14).
\label{ratio}
\end{eqnarray}

As seen, the values of these ratios are very close to each other;
the largest difference between $-2.1790$ and $-2.1656$ is only about 0.6\%.
This is not surprising because the electronic matrix elements $\langle \gamma J ||T|| \gamma J \rangle$ are practically the same
for both $^{233}$U and $^{235}$U. A small difference in nuclear radii affects the ratio of these matrix elements very little.

Using the average of the three ratios given in~\eref{ratio}, $-2.1722$, and
assuming that this is about the same also for the hfs constants of the $^6\!L_{11/2}$ and $^6\!K_{9/2}$ states, we found the respective
values $A(^6\!L_{11/2})$ and $A(^6\!K_{9/2})$, presented in the last column of \tref{hfs}.

%At the same time, any direct measurement of these hfs constants would be very helpful and useful.
%Taking into account that the calculations of these hfs constants were done in the same way,
%it is difficult to explain the reason for this deterioration in accuracy.
%----------------
\section{Pure CI calculations}
\label{CI}
%----------------
One of the reasons for the low accuracy of the hfs constants of the excited states of U$^+$ is a poor initial approximation.
Indeed, the initial DHF self-consistency procedure was done for the core $[1s^2, ...,6s^2,6p^6]$ electrons. Then the valence orbitals were
constructed in the frozen-core approximation, i.e., in the field of the sixfold ionized neutral atom, whereas we would be interested in constructing
orbitals for the neutral atom.
In this section, we discuss another (pure CI) method of calculation that allows us to use a much better initial approximation.
The role of the core-valence correlations can be estimated by successively adding the core shells ($6p$, $6s$, etc.) to the valence field and
by calculating the hfs constants for these extended CI spaces.
Such a method is similar to the multiconfiguration Dirac-Hartree-Fock (MCDHF) variational approach~\cite{FroGodBra16}.
In particular, in this review paper the authors discuss an optimization strategy for choosing orbitals that should
be involved in the MCDHF process.

Again, at first U$^+$ and U were considered as the atomic systems with five and six valence electrons, respectively, above
the closed core $[1s^2,\,.\,.\,.\,,6p^6]$.
But this time, we solved the DHF equations in the $V^N$ approximation for both atomic systems, i.e.,
the initial self-consistency procedure was carried out for the $[1s^2,\,.\,.\,.\,,6p^6,5f^3\,6d\,7s]$ configuration for U$^+$ and
for the $[1s^2,\,.\,.\,.\,,6p^6,5f^3\,6d\,7s^2]$ configuration for U. For U$^+$, all electrons were frozen, the electron from the
$6d$ shell was moved to the $7p$ shell, and the $7p_{{1/2},{3/2}}$ orbitals were constructed in the frozen-core potential.
The remaining virtual orbitals were formed using a recurrent procedure described in Refs.~\cite{KozPorFla96,KozPorSaf15}.
%when the large component of the radial Dirac bispinor, $f_{n'l'j'}$, was obtained from a previously constructed function $f_{nlj}$ by
%multiplying it by $r^{l' - l}\, \sin(kr)$, where $l'$ and $l$ are the orbital quantum numbers of the new and old orbitals ($l' \geq l$) and
%the coefficient $k$ is determined by the properties of the radial grid. The small component $g_{n'l'j'}$ was found from the kinetic balance condition.
% \begin{equation}\label{kbal}
% g_{nlj} =\frac{\bm \sigma \bm p}{2mc} f_{nlj}\,,
% \end{equation}
%where $\bm\sigma$ are the Pauli matrices, ${\bm p}$ and $m$ are the electron momentum and mass, and $c$ is the speed of light.
The newly constructed orbitals were then orthonormalized with respect to the orbitals of the same symmetry.

For both atomic systems, the basis sets included in total five partial waves ($l_{\rm max} = 4$) and the orbitals with the principal quantum number
$n$ up to 20. We included the Breit interaction on the same footing as the Coulomb interaction when constructing the basis set.
%QED corrections were also included following Refs.~\cite{ShaTupYer13,TupKozSaf16}.

We note that with such a construction of the basis set the number of electrons included in the DHF self-consistent procedure differs
from the number of the core electrons. As discussed earlier, we cannot use the CI+all-order method with such a starting potential.
So, we apply the pure CI method, accounting for explicitly only the valence-valence correlations but considering more and more electrons
as the valence electrons in successive approximations.
%-----------------------------------------------------------------------------------
\subsection{The hfs constant $A$ of the $(5f^3\,6d7s\;^6\!L_{11/2})$ state for U$^+$}
\label{HFSU+}
%-----------------------------------------------------------------------------------
Our goal is to calculate the hfs constant $A$ of the first excited state $(5f^3\, 6d7s\;\;^6\!L_{11/2})$ (for which the experimental
value is known) in the framework of the pure CI method.
At first, we consider U$^+$ as the ion with five valence electrons and, respectively, we do five-electron CI
(in the following, we designate it as 5$e$ CI).

We construct the sets of configurations by allowing single and double excitations of electrons from the two main configurations $5f^3\,7s^2$ and
$5f^3\,7s6d$ to higher-lying orbitals. We successively increased the configuration space, following the change of the constant $A$, until it stops changing.
%At this stage, we were able to saturate CI, successively increasing the number of configurations until the constant $A$ stops changing.
The largest set of configurations was constructed by single and double excitations to $[13spdf]$.
%and included 2.2 million determinants.
The results obtained for 5$e$ CI for different sets of configurations are given in the first row of \tref{HFS_U+}. As seen, practically the same values were found for the $[10,11,12,13spdf]$ sets of configurations.

As a next step, we estimated the role of the core-valence correlations by successively including the $6p$, $6s$, and $5d$ shells in the valence field
and carrying out calculations in the framework of 11-, 13-, and 23-electron CI, respectively. The results obtained for different sets of configurations
are presented in rows 2-4 in~\tref{HFS_U+}.

Finally, we included the $5p$ and $5s$ shells in the valence field and carried calculations in the framework of $29e$ CI and $31e$ CI. The largest set of configurations for $29e$ CI was $[9spdf]$ while for $31e$ CI it was $[8spdf]$. Due to a very large number of determinants we are unable to
carry out calculation for $[9spdf]$ in the framework of $31e$ CI, but we can do the following estimate.
The contribution from the excitations to the $9spdf$ shells for $29e$ CI can be determined as $A([9spdf])-A([8spdf]) = -29\, {\rm MHz}$.
Assuming that the contribution from the excitations to the $9spdf$ shells for $31e$ CI is comparable, we would obtain for $31e$ CI
the value of the constant $A[9spdf] \simeq -313$ MHz which is close to the value obtained for $23e$ CI. Thus,
when we include the $5p$ and $5s$ shells in the CI space, their contributions to the constant $A$ tend to cancel each other.

To find out the role of excitations to the $g$ shells we compared the results for $[8spdf]$ and $[8spdf6g]$. They are given in \tref{HFS_U+}
in columns 3 and 4. For different sets of configurations we observe that the excitations to the $5g$ and $6g$ shells change the hfs constant
at the level of few MHz. For the $29e$ CI the difference $A([8spdf6g])-A([8spdf]) = -1\, {\rm MHz}$. We also checked that the excitations
to the $7g-10g$ shells, practically, do not play any role. They lead to a change of the constant $A$ by less than 1 MHz.

%#####################################################################################################################################
\begin{table*}[tp]
\caption{Magnetic dipole hfs constants $A$ (in MHz) of the $5f^3\, 6d7s\;\;^6\!L_{11/2}$ state of U$^+$, obtained in different approximations
with $\mu = 0.59\,\mu_N$, are presented. The experimental value is given in the last column. The final theoretical value is given in the
row labeled ``Final''. Uncertainties are given in parentheses.}
\label{HFS_U+}%
\begin{ruledtabular}
\begin{tabular}{cccccccccc}
         &$[7spdf]$& $[8spdf]$ & $[8spdf6g]$ & $[9spdf]$ & $[10spdf]$ & $[11spdf]$ & $[12spdf]$ & $[13spdf]$ & Experiment~\cite{Thi21}\\
\hline      \\[-0.6pc]
$5e$ CI  &  -234   &   -264    &    -272     &   -264    &   -263     &   -263     &    -263    &    -263    & -$301(6)$  \\[0.2pc]

$11e$ CI &  -237   &   -283    &    -283     &   -289    &   -286     &   -285     &    -286    &    -287    &            \\[0.2pc]

$13e$ CI &  -242   &   -297    &    -292     &   -315    &   -310     &   -301     &    -296    &    -296    &            \\[0.2pc]

$23e$ CI &  -249.6 &   -291    &    -288     &   -313    &   -311     &            &            &            &            \\[0.2pc]

$29e$ CI &  -250.9 &   -294    &    -295     &   -323    &            &            &            &            &            \\[0.2pc]

$31e$ CI &  -250.6 &   -284    &             &           &            &            &            &            &            \\[0.2pc]

 Final   &         &           &             &           & -$298(10)$ &            &            &            &
\end{tabular}
\end{ruledtabular}
\end{table*}

As we observed for $11e$ and $13e$ CIs, the CI spaces were saturated for the $[13spdf]$ set of configurations.
We consider the value $-296$ MHz, obtained for $13e$ CI for the $[13spdf]$ set of configuration, as closest to the final result.
To estimate corrections to this value due to contributions of the deeper-lying core shells,
we added the difference $A(23e\, {\rm CI}) - A(13e\, {\rm CI}) = -1\, {\rm MHz}$, found for the $[10spdf]$ set
of configurations, and also the difference $A([8spdf6g]) - A([8spdf]) = -1\, {\rm MHz}$ found for $29e$ CI, to $-296$ MHz,
arriving at $-298$ MHz. As we already mentioned, the excitations from the $5p$ and $5s$ shells give contributions to the constant $A$
tending to cancel each other.

We were unable to include into consideration the excitations from the $1s,...,4d$ core shells. But they are lying already sufficiently deep
(e.g., the DHF energy of the $4d_{5/2}$ orbital is $-28$ a.u.).
We believe that the total contribution from excitations from these shells
to the final value of the hfs constant $A$ does not exceed the difference between the results, obtained
for $[9spdf]$ for $29e$ CI and $23e$ CI, equal to 10 MHz. We consider this value as the uncertainty of our calculation.

Using the theoretical and experimental values for $A(^6\!L_{11/2})$, $-298(10)$ and $-301(6)$ MHz~\cite{Thi21}, respectively, we are able to extract
the nuclear magnetic moment of $^{233}$U$^+$ to be $\mu/\mu_N \approx 0.596(24)$. We note that this value is somewhat inconclusive because we used
only one hfs constant. Further experimental measurements of other hfs constants of the low-lying states are needed.
By performing similar calculations and analyses for these constants, we would be able to refine the value of the magnetic moment. Besides that, it would
allow us to investigate the problem of the Bohr-Weisskopf effect of the finite nuclear magnetization, which is omitted in the present calculation due to the lack of further experimental values that will allow its extraction (see Ref.~\cite{PorSafKoz21} for detail). We note that
if the nuclear magnetic moment of $^{233}$U is known with good accuracy and also knowing the highly accurate value of
the ratio of the nuclear magnetic moments $\mu(^{233}{\rm U})/\mu(^{235}{\rm U}) = 1.5604(14)$~\cite{GanZemKul90,Sto05}
we are able to determine the nuclear magnetic moment of $^{235}$U.
%--------------------------------------------------------------
\subsection{hfs constants $A$ of the low-lying states of U}
\label{HFSU}
%--------------------------------------------------------------
An accurate calculation of different properties of the neutral uranium is generally a more complicated task due to an extra valence electron in comparison to U$^+$.
We consider U as the atom with six valence electrons and, respectively, we do the six-electron CI, constructing the configurations by allowing
single and double excitations from the two main configurations $5f^3\,7s^2\,6d$ and $5f^3\,7s\,6d^2$ to higher-lying orbitals.
The results are presented in \tref{HFS_U}. As seen from the table, the results obtained for the $[10spd8fg]$ and $[11sp10dfg]$
sets of configurations are very close to each other for all four considered states. So we can say that the saturation of $6e$ CI is achieved.
The values of the hfs constants are in reasonable agreement with the experimental data.
Unfortunately, a reliable calculation of the hfs constants in the framework of 12$e$ CI (when the $6p$ shell was included in the valence field)
proved impossible even with modern computer capabilities. Even for a rather small $[7spdfg]$ set of configurations, the CI space consisted
of $84 \times 10^6$ determinants.
%#####################################################################################################################################
\begin{table*}[tp]
\caption{Magnetic dipole hfs constants $A$ (in MHz) of the four lowest-lying states of $^{233}$U obtained (with $\mu = 0.59\,\mu_N$)
in the framework of 6$e$ CI in different approximations. The experimental values are given in the last column.}
\label{HFS_U}%
\begin{ruledtabular}
\begin{tabular}{ccccccc}
         &                             & $[7spdfg]$ & $[8spdfg]$ & $[10spd8fg]$ & $[11sp10dfg]$ & Experiment \\
\hline      \\[-0.6pc]
$6e$ CI  & $A(5f^3\, 6d 7s^2~^5\!L_6)$ &    134     &    107     &     114      &     115       &  $131.56(10)^{\rm a}$ \\[0.1pc]
         & $A(5f^3\, 6d 7s^2~^5\!L_7)$ &    136     &    125     &     126      &     126       &  $122^{\rm b}$        \\[0.1pc]
         & $A(5f^3\, 6d 7s^2~^5\!K_6)$ &     40     &    123     &     123      &     124       &  $128^{\rm b}$        \\[0.1pc]
         & $A(5f^3\, 6d^2 7s~^7\!M_6)$ &    -39     &   -135     &    -144      &    -144       & -$147^{\rm b}$        %\\[0.3pc]
%$12e$ CI & $A(5f^3 6d 7s^2\,^5\!L_6)$  &    124     &            &              &               &                       \\[0.1pc]
%         & $A(5f^3 6d 7s^2\,^5\!L_7)$  &    135     &            &              &               &                       \\[0.1pc]
%         & $A(5f^3 6d 7s^2\,^5\!K_6)$  &    134     &            &              &               &                       \\[0.1pc]
%         & $A(5f^3 6d^2 7s\,^7\!M_6)$  &   -125     &            &              &               &
\end{tabular}
\end{ruledtabular}
\begin{flushleft}
$^{\rm a}$Reference~\cite{GanZemKol97}; \\
$^{\rm b}$Recalculated from the results presented in Ref.~\cite{AvrGinPet94}.
\end{flushleft}
\end{table*}
%#####################################################################################################################################
% ------------------
\section{Conclusion}
% ------------------
In summary, we carried out calculations of the energy levels and hfs constants for a number of the low-lying states of U$^+$ and U in the framework
of the CI+all-order and pure CI methods. Using the calculated and experimental results of the magnetic dipole hfs constant $A(^6\!L_{11/2})$
of U$^+$, we extracted the value of the magnetic dipole moment $\mu_N$ of the nucleus of $^{233}$U. Further experimental measurements of
other magnetic dipole hfs constants of the low-lying states and accurate theoretical calculations are needed to confirm the value
of $\mu_N$ obtained in this paper.

In the framework of the CI+all-order method, we calculated the corrections to the hfs operator in the second order of the perturbation theory.
For the hfs constants of the excited states of U$^+$, these corrections are large, leading to large uncertainties of these values.
To be more certain about these quantities, their calculation in all orders of the perturbation theory is required.

In terms of the experiment, a measurement of the magnetic dipole hfs constant $A$ of the ground state $(5f^3\,7s^2\;^4\!I_{9/2})$ of $^{233}$U$^+$,
as well as a direct measurement of the hfs constants of low-lying states of the neutral $^{233}$U, such as $(5f^3\,6d 7s^2\;^5\!L_7)$ and
$(5f^3\, 6d 7s^2\;^5\!K_6)$ would be very useful. This would contribute to further development of the theory, and a better understanding of the possibilities of modern methods of high-precision calculations, and would allow an accurate extraction of U nuclear moments.
% -----------------------
%\section{Acknowledgments}
% -----------------------

This work is a part of the ``Thorium Nuclear Clock'' project that has received funding from the European Research Council under
the European Union's Horizon 2020 research and innovation program (Grant No. 856415).
This work was supported in part by NSF Grant No.~PHY-2012068 and
through the use of UDEL Caviness and DARWIN computing systems: DARWIN - A Resource for Computational and Data-intensive Research at the University of Delaware and in the Delaware Region, Rudolf Eigenmann, Benjamin E. Bagozzi, Arthi Jayaraman, William Totten, and Cathy H. Wu, University of Delaware, 2021, URL: https://udspace.udel.edu/handle/19716/29071.


%\bibliography{U}

\end{document}